\documentclass[twocolumn,pra,aps,showpacs]{revtex4}
\usepackage[dvips]{graphicx}
\usepackage{graphicx}% Include figure files
\usepackage{dcolumn}% Align table columns on decimal point
\usepackage{bm}% bold math
\begin{document}

\title{Voltage-tunable lateral shifts of ballistic electrons in
semiconductor quantum slabs}

\author{Xi Chen$^{1}$\footnote{Email address: xchen@shu.edu.cn}}

\author{Yue Ban$^{1}$}

\author{Chun-Fang Li$^{1}$\footnote{Email address: cfli@shu.edu.cn}}

\affiliation{$^{1}$ Department of Physics, Shanghai University,
Shanghai 200444 , People's Republic of China}

\begin{abstract}

It is investigated that the lateral shifts of the ballistic
electrons transmitted through a semiconductor quantum slabs can be
negative as well as positive, which are analogous to the anomalous
lateral shifts of the transmitted light beam through a dielectric
slab. The necessary condition for the shift to be negative is
advanced. It is shown that the lateral shifts depend not only on the
structure parameters of semiconductor quantum slab, but also on the
incidence angle and the incident energy. Numerical calculations
further indicate that the lateral shifts can be tuned from negative
to positive by the external applied electric field. The
voltage-tunable lateral shifts may lead to potential applications in
quantum electronic devices.



\end{abstract}

\maketitle
%----------------------------------------------------------------

\section{Introduction}

The analogies between phenomena occurring in two different physical
systems open a route to find new effects or to translate solution on
techniques or devices, and quite often help to understand both
systems better \cite{Dragoman}. In particular, electronic analogues
of many optical behaviors such as reflection, refraction
\cite{Gaylord,Gaylord-GB}, focusing \cite{Houten,Spector-S-K,Sivan},
collimation \cite{Molenkamp}, and interference
\cite{Yacoby,Liang,Ji} have been achieved in two-dimensional
electron gas (2DEG) enabling the systems as a basic platform to
study foundation problems in quantum mechanics \cite{Buks,Chang} as
well as quantum information processing \cite{Beenakker-EKV}. The
close relation between optics and electronics result from the fact
that the electrons act as wave due to the ballistic transport
properties of a highly mobility 2DEG created in semiconductor
heterostructures \cite{Palevski}. As a result, there is growing
interest in the design and development of devices based on electron
wave propagation, which have given rise to a research field
described as electron wave optics
\cite{Gaylord-Brennan-Gaylord-G-H,Dragoman-D,Datta}.

Among those electric-optic analogies, the electric analogue of
Goos-H\"{a}nchen (GH) effect have attracted much attention in recent
years \cite{Wilson-G-G,Zhao,Beenakker,Chen,Chen-PRB}, because of its
practical significance in design of many new quantum electron
devices. In 1993, Gaylord {\it et al.} \cite{Wilson-G-G} once showed
that the ballistic electrons totally reflected from a
potential-energy/effective-mass interface experience a lateral shift
from the position predicted by electron wave optics. This phenomenon
is known as GH effect in optics \cite{Goos}. Very recently, Zhao and
Yelin \cite{Zhao} have shown that, based on the electronic
counterpart to the trapped rainbow effect in optics
\cite{Tsakmakidis}, the interplay GH effect and negative refraction
\cite{Cheianov} in graphene leads to the coherent graphene devices,
such as movable mirrors, buffers and memories. Beenakker \textit{et
al.} \cite{Beenakker} have further found that the GH effect at a
$n$-$p$ interface in graphene doubles the degeneracy of the lowest
propagating mode, which can be observed as a stepwise increase by
$8e^2/h$ of the conductance with increasing channel width. As a
matter of fact, the GH shifts for matter wave \cite{Renard,Carter}
and even relativistic Dirac electrons \cite{Miller,Fradkin} have
been considered previously. However, it is shown that the electric
GH shift in total reflection is only about the order of the electron
wavelength. The smallness of the GH shift has impeded its direct
measurement and applications. In previous paper, we found that the
lateral shifts of ballistic electrons in a semiconductor quantum
barrier can be enhanced by transmission resonance, and are always
positive \cite{Chen}. It is further shown that the tunable lateral
shifts in magnetic-electric nanostructure have some significant
applications in spin filter and spin beam splitter \cite{Chen-PRB}.

In present paper, we will investigate that the lateral shift of
ballistic electrons transmitted through a semiconductor quantum well
\cite{Gaylord-GB,Wilson-G-G}, acting as electron wave slab, can be
negative as well as positive. The necessary condition for the
lateral shift to be negative is given. The negative lateral shift
presented here is similar to but different from a phenomenon of
electronic negative refraction \cite{Cheianov,Dragoman-JAP}.
Actually, the negative lateral shift is the electronic analogue to
the negative lateral shift of a light beam transmitted through a
dielectric slab \cite{Li}. More importantly, it is found that the
lateral shifts in a fixed semiconductor heterostructure can be
modulated from negative to positive by the external applied electric
field. This leads to some interesting applications in quantum
electronic devices such as beam spatial modulation, beam splitter,
and wave vector or energy filter.

\section{Lateral shifts in semiconductor quantum slab}
\label{Sec II}

\begin{figure}[]
\scalebox{0.72}[0.72]{\includegraphics{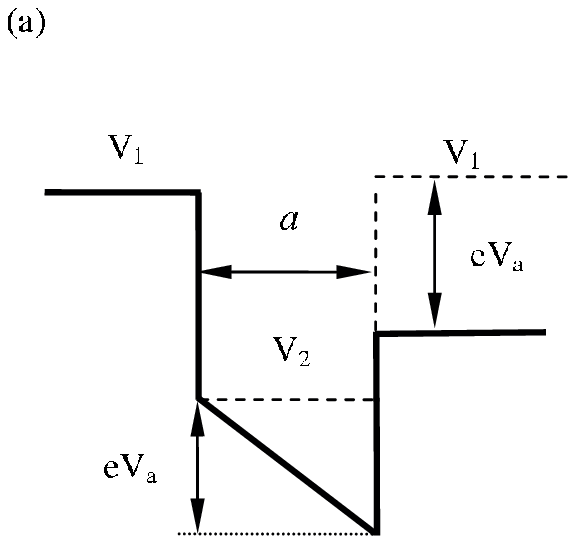}}
\scalebox{0.72}[0.72]{\includegraphics{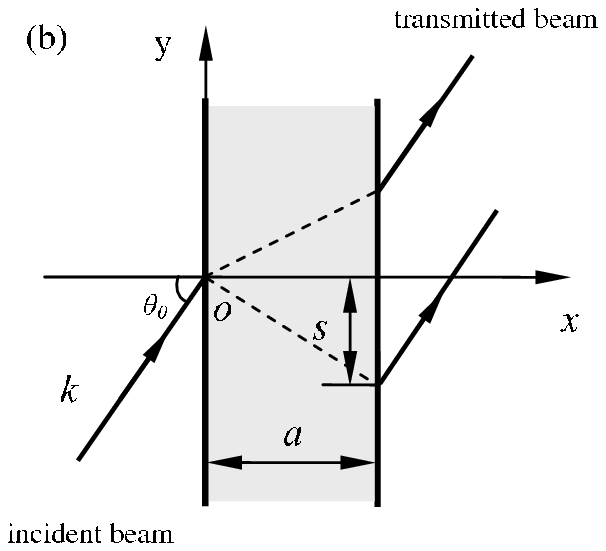}}
\caption{Schematic diagram of positive and negative lateral shifts
of ballistic electrons propagating obliquely through a quantum slab,
corresponding to a two-dimensional semiconductor potential well
under external applied electric field.} \label{fig.1}
\end{figure}

Consider ballistic electrons with the energy $E$ is incident
obliquely upon a two-dimensional semiconductor quantum well,
extending from $0$ to $a$, as shown in Fig. \ref{fig.1}, where
$V_2<V_1$, and $V_2$ and $V_1$ are the depth of potential well and
the values of the potential energies on its two sides, respectively.
The corresponding electron effective masses are $m^*_1$ and $m^*_2$.
The wave function of an incident electron beam can be expressed in
terms of plane wave components as $
\Psi_{i}(x,y)=\int_{-k}^{k}A(k_y)\exp[i(k_x x+k_y y)]dk_y, $ where
$\vec{k}=(k_x, k_y)=(k \cos\theta, k \sin\theta)$,
$k=[2m^*_1(E-V_1)]^{1/2}/\hbar$, $\hbar$ is defined by Plank
constant $h$ divided by $2\pi$, $\theta$ stands for the incident
angle of the plane wave component under consideration, and $A(k_y)$
is the amplitude angular-spectrum distribution of electron beam. If
the angular spectral distribution $A(k_y)$ is sharp enough, the wave
function can be rewritten as
\begin{equation}
\Psi_{i}(x,y)=\int_{-\infty}^{\infty}A(k_y)\exp[i(k_x x+k_y y)]dk_y.
\end{equation}
For a Gaussian-shaped incident beam whose peak is assumed to be
located at $x=0$,
\begin{equation}
\label{incident beam}
\Psi_{i}(x=0,y)=\exp{\left(-\frac{y^2}{2w^2_y}\right)}\exp{(i
k_{y0}y)},
\end{equation}
its angular spectral distribution is also a Gaussian function,
$A(k_y)=w_y \exp[-(w_y^2/2)(k_y-k_{y0})^2]$, around its central
$k_{y0}=k \sin \theta_0$, $w_y=w_0/\cos\theta_0$, $w_0$ is the width
of the beam at waist. For a incident plane wave component
$\exp{[i(k_x x+k_y y)]}$, letting be $T(k_y)\exp{\{i[k_x (x-a)+k_y
y]\}}$ the corresponding transmitted plane wave component, the
transmitted beam at $x=a$ can be expressed by,
\begin{equation}
\label{transmission} \Psi_t (a,y)= \int_{-\infty}^{\infty}T(k_y)
A(k_y)\exp{(i k_y y)}dk_y,
\end{equation}
where the transmission coefficient $T(k_y)=e^{i \phi}/g$ is
determined by the following complex number,
$$
ge^{i \phi}=\cos k'_x a+
\frac{i}{2}\left(\frac{m^*_2}{m^*_1}\frac{k_x}{k'_x}+
\frac{m^*_1}{m^*_2}\frac{k'_x}{k_x}\right) \sin k'_x a,
$$
so that the phase shift can be expressed by
\begin{equation}
\label{phaseshift} \tan \phi= \frac{1}{2}\left(
\frac{m^*_2}{m^*_1}\frac{k_x}{k'_x}+
\frac{m^*_1}{m^*_2}\frac{k'_x}{k_x} \right) \tan k'_x a,
\end{equation}
where $k'_x=k'\cos \theta'$, $k'=[2m^*_2(E-V_2)]^{1/2}/\hbar$, and
$\theta'$ is determined by Snell's law for electron wave, $
\sin\theta'/\sin \theta= [m^*_1(E-V_1)/m^*_2(E-V_2)]^{1/2}$. For a
well collimated beam, the lateral shift of the ballistic electrons
in transmission the quantum slab is defined as $-d\phi/dk_{y0}$
\cite{Wilson-G-G,Chen}, according to the stationary phase method
\cite{Bohm}, and is thus given by
\begin{widetext}
\begin{eqnarray}
\label{lateral displacement} s &=& \frac{s_{g}}{2 g^2_0} \left[
\left(\frac{m^*_2}{m^*_1}\frac{k_{x0}}{k'_{x0}}+
\frac{m^*_1}{m^*_2}\frac{k'_{x0}}{k_{x0}}\right)- %\nonumber \\ &&
\left(1-\frac{k'^2_{x0}}{k_{x0}^2} \right)
\left(\frac{m^*_2}{m^*_1}\frac{k_{x0}}{k'_{x0}}-
\frac{m^*_1}{m^*_2}\frac{k'_{x0}}{k_{x0}}\right) \frac{\sin 2k'_{x0}
a }{2k'_{x0} a} \right],
\end{eqnarray}
\end{widetext}
where $k'_{x0}=k' \cos \theta'_0$, $\theta'_0$ is determined by
Snell's law for electron wave, $\sin\theta'_0/\sin \theta_0=
[m^*_1(E-V_1)/m^*_2(E-V_2)]^{1/2}$, $s_{g}= a \cdot \tan \theta'_0$
is the geometrical shift predicted by electron wave optics, and
$|T_0|^2=1/g^2_0$ is the transmission probability, which is closely
related to the measurable ballistic conductance $G$, according to
the well-known Landauer-B\"{u}ttiker formula \cite{Buttiker}. It is
noted that the subscript $0$ in this paper denotes the values taken
at $k_y=k_{y0}$, namely $\theta=\theta_0$. It is clear that the
lateral shift $s$ is modulated by the transmission probability as is
indicated in Eq. (\ref{lateral displacement}), thus is in general
different from $s_{g}$. More interestingly, when the necessary
condition
\begin{equation}
\label{inequality} \frac{m^*_2}{m^*_1}\frac{k_{x0}}{k'_{x0}}+
\frac{m^*_1}{m^*_2}\frac{k'_{x0}}{k_{x0}}<
\left(1-\frac{k'^2_{x0}}{k_{x0}^2} \right)
\left(\frac{m^*_2}{m^*_1}\frac{k_{x0}}{k'_{x0}}-
\frac{m^*_1}{m^*_2}\frac{k'_{x0}}{k_{x0}}\right), \end{equation}
since $\sin 2k'_{x0} a /2k'_{x0} a\leq 1$, the lateral shift can be
negative when the restriction to the incidence angle $\theta_0$,
\begin{equation}
\label{restriction to incident angle} \cos \theta_0 <
\left[\frac{m^*_2(E-V_2)/m^*_1(E-V_1)-1}{1+(m^*_2/m^*_1)^2}\right]^{1/2}
\equiv \cos \theta_t,
\end{equation}
is satisfied. This means that if $\theta_0$ is larger than the
threshold angle $\theta_t$, one can always find a thickness $a$ of
the quantum wells where the lateral shift of the transmitted beam is
negative. Moreover, it is clearly seen that when the incident energy
is in the region of
\begin{equation}
V_1<E<E_c\equiv\frac{m^*_1 V_1-m^*_2 V_2}{m^*_1-m^*_2},
\end{equation}
that is, $m^*_2(E-V_2)/m^*_1(E-V_1)>1$, the necessary condition
(\ref{restriction to incident angle}) is satisfied, so that the
lateral shift of ballistic electrons in the quantum slab can be
negative as well as positive. On the contrary, the lateral shifts in
the case of $E>E_c$ are always positive, since the necessary
condition (\ref{restriction to incident angle}) is invalid. In fact,
when the semiconductor quantum barrier ($V_2>V_1$) is considered,
the critical angle for total reflection is defined as,
\begin{eqnarray}
\theta_{c}=  \sin^{-1} \sqrt{\frac{m_2^*(E-V_2)}{m_1^*(E-V_1)}}, &
V_2< E< E_c,
   \end{eqnarray}
Thus, the lateral shifts are always positive, as discussed in Ref.
\cite{Chen}. More interestingly, when $E$ is larger than the
critical energy $E_c$, we have critical angle $\pi/2$, that is, the
ballistic electrons can traverse through the quantum barrier at any
incidence angle. In this case, the quantum barrier behaves as
quantum slab for $V_1<E<E_c$ \cite{Wilson-G-G}, thus the lateral
shifts can be negative under the condition (\ref{restriction to
incident angle}).

\begin{figure}[]
\scalebox{0.40}[0.40]{
  \includegraphics{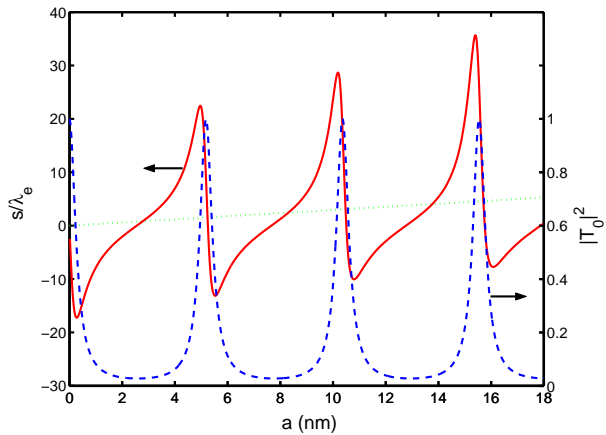}}
\caption{(Color online) Dependence of the lateral shift (solid
curve) and transmission probability (dashed curve) on the width $a$
of the quantum slab, where $V_1=231.92 meV$, $V_2=0$,
$m^*_1=0.092m_0$, $m^*_2=0.067 m_0$, $E=300meV (E_c=856 meV)$, and
$\theta_0= 80^{\circ}$ ($\theta_t =46.1^{\circ}$). The dotted curve
represents the lateral displacement predicted from electron wave
optics, $s_{g}$.} \label{fig.2}
\end{figure}

To illustrate the negative lateral shifts of ballistic electrons,
the example semiconductor quantum well consisting of
$\mbox{Ga}_{1-\chi}\mbox{Al}_{\chi}\mbox{As/GaAs/}\mbox{Ga}_{1-\chi}\mbox{Al}_{\chi}\mbox{As}$.
The potential energies in the three regions of the quantum well are
given by the conduction band edge as $V_1=A\chi$, $V_2=0$ and
$A=773.1 meV$ (in which the conduction band discontinuity has been
taken to be $60\%$ of the energy gap change). The corresponding
effective electron mass is $m^*_1=(B+C\chi)m_0$, where $m_0$ is the
free electron mass, $B=0.067$ and $C=0.083$ \cite{Wilson-G-G}. Fig.
\ref{fig.2} shows the typical dependence of the lateral shift (solid
curve) and transmission probability (dashed curve) on the width $a$
in semiconductor quantum well
$\mbox{Ga}_{0.7}\mbox{Al}_{0.3}\mbox{As/GaAs/}\mbox{Ga}_{0.7}\mbox{Al}_{0.3}\mbox{As}$,
where $V_1=231.92 meV$, $V_2=0$, $m^*_1=0.092m_0$, and
$m^*_2=0.067m_0$, the incidence energy is $E=300meV (E_c=856 meV)$ ,
the incidence angle is $80^{\circ}$, $s$ is in units of the wave
length of free electron, $\lambda_e= h/\sqrt{2m_0E}$, and the dotted
curve represents the lateral displacement predicted from electron
wave optics, $s_{g}$. It is clear that the lateral shift $s$ is
quite different from the lateral displacement predicted from
electron wave optics, and can be negative. The negative shift is
similar to but different from the negative refraction of ballistic
electrons in graphene \cite{Cheianov} or metamaterial
\cite{Dragoman-JAP}. This phenomenon results from the destructive
and constructive interference of the electron waves due to the
multiple reflections and transmissions inside the slab. As a matter
of fact, the inequality (\ref{inequality}) is required for the
lateral shift to be negative. Since the function $\sin 2k'_{x0}
a/2k'_{x0} a$ decreases rapidly with increasing $k'_{x0} a$, the
width of the potential well should be of the order of $\pi/k'_{x0}$,
that is, the order of the wavelength $\lambda_e$, so as to make the
negative lateral shift significantly large. As a result, the
necessary condition for the width of quantum slab \cite{Chen}
\begin{equation}
\label{restriction-a} a \ll a_c\equiv \frac{w_0}{2 \cos \theta_{0}
\tan \theta'_{0}},
\end{equation}
holds for the incident beam, the whole transmitted beam will
maintain the shape of incident beam with negative lateral shift,
that is to say, the stationary phase method is valid when the
condition (\ref{restriction-a}) is satisfied.

\begin{figure}[]
\scalebox{0.40}[0.40]{
  \includegraphics{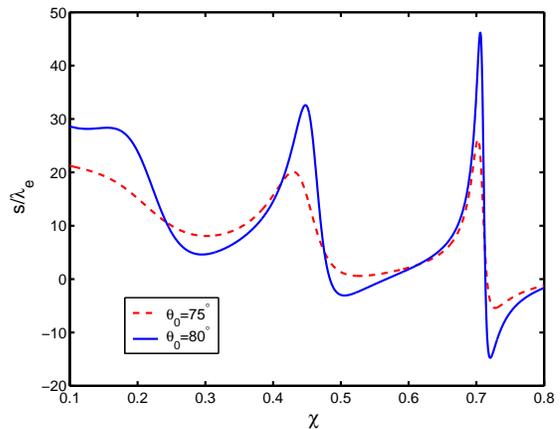}}
\caption{(Color online) Dependence of the lateral shift on the
physical parameter $\chi$ of the semiconductor quantum slab at
different incidence angles, where $V_1=A\chi$, $m^*_1=(B+C\chi)m_0$,
$E=600meV$, $a=10 nm$, and other physical parameters are the same as
in Fig. \ref{fig.2}.} \label{fig.3}
\end{figure}

It should be pointed out that the physical structure of
semiconductor quantum well has great impact on the lateral shift.
Fig. \ref{fig.3} shows the dependence of the lateral shift on the
parameter $\chi$ of the semiconductor quantum slab at different
incidence angles, where $V_1=A\chi$, $m^*_1=(B+C\chi)m_0$,
$E=600meV$, $a=10 nm$, and other physical parameters are the same as
in Fig. \ref{fig.2}. It is shown that the absolute value of lateral
shift can be enhanced by increasing the physical parameter $\chi$,
which corresponds to the increase of the effective mass and
potential energies in the two sides of quantum slab. In addition,
Eq. (\ref{lateral displacement}) also indicates that the negative
lateral shift also depends on the angle of incidence and the
incident energy. Thus, the spatial location of the transmitted
ballistic electrons can be altered by the negative and positive
lateral shifts which can be easily controlled by the incidence angle
or the incidence energy. In what as follows we will emphasize the
modulation of the lateral shifts in a fixed semiconductor quantum
structure by external electric field.

\section{Modulation by external applied electric field}

In this section, we will focus on the control of the lateral shifts,
taking account into imposing external electric field on the
semiconductor quantum slab as shown in Fig. \ref{fig.1}(a).
\begin{figure}[]
\scalebox{0.40}[0.40]{
  \includegraphics{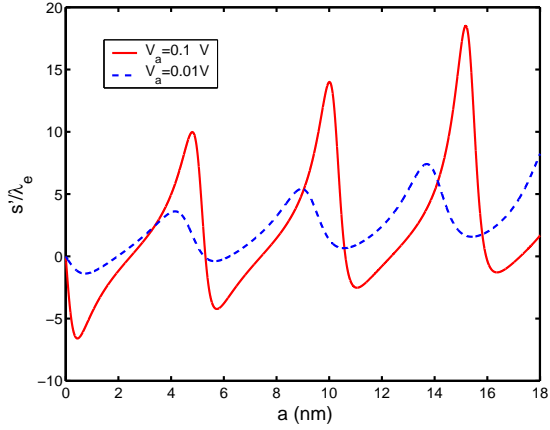}}
\caption{(Color online) Dependence of the lateral shift on the width
$a$ of the quantum slab, where $V_a =0.1 V$ (solid curve) and $V_a
=0.01 V$ (dashed curve), and the other physical parameters are the
same as in Fig. \ref{fig.2}.} \label{fig.4}
\end{figure}
For a plane wave component $\exp{[i(k_x x+k_y y)]}$ of the incident
ballistic electrons, the corresponding transmitted plane wave is
given by $T'(k_y)\exp{\{i[k'_x (x-a)+k_y y]\}}$ in this case, where
$k'=[2m^*_1(E-V_1+eV_a)]^{1/2}/\hbar$. Because the system is
translationally invariant along the $y$ direction, the wave function
in the region of the quantum well under applied electric field can
be expressed as $\Psi (x,y)=\psi(x)e^{ik_y y}$, where the
longitudinal wave packet is determined by,
\begin{equation}
\label{Shrodinger equation} \frac{\partial^2 \psi(x)}{{\partial
x^2}} +\frac{2m^*_2}{\hbar^2}\left(E_x-V_2+
\frac{eV_ax}{a}\right)\psi(x) = 0,
\end{equation}
where $E_x=\hbar^2 k^2_x/2 m^*_2$ is the longitudinal energy and
$V_a$ is the applied biased voltage. The solutions of the above
Schr\"{o}dinger equation are the well-known linearly independent
Airy functions $Ai(\eta)$ and $Bi(\eta)$ \cite{Mathematics}, that is
$\psi(x) = M Ai(\eta) + N Bi(\eta)$, where
$$\eta = \left(-\frac{{2
m^*_1 e V_a }}{{a \hbar^2}}\right)^{1/3} \left[\frac{a}{{e
V_a}}(E_x-V_2)+x \right].$$ Combining the boundary conditions, we
can finally obtain the transmission coefficient $T'(k_y)=e^{i
\phi'}/g'$ determined by the following complex number
$$
g'e^{i\phi'}=\frac{1}{2\gamma}\left[\left(\mu + \frac{k'_{x}}{k_{x}
}\nu\right)+i\left(\frac{m^*_2}{m^*_1}\frac{k'_x}{\kappa}\alpha+\frac{m^*_1}{m^*_2}\frac{\kappa}{k_x}\beta\right)\right],
$$
so that
\begin{equation}
\tan \phi' =
\left(\frac{m^*_2}{m^*_1}\frac{k'_x}{\kappa}\alpha+\frac{m^*_1}{m^*_2}\frac{\kappa}{k_x}\beta\right)/\left(\mu
+ \frac{k'_{x}}{k_{x} }\nu\right),
\end{equation}
where
\begin{eqnarray*}
\alpha &=& Ai(0)Bi(a)-Ai(a)Bi(0),\\
\beta &=& Ai'(0)Bi'(a)-Ai'(a)Bi'(0),\\
\mu &=& Ai(0)Bi'(a)-Ai'(a)Bi(0),\\
\nu &=& Ai(a)Bi'(0)-Ai'(0)Bi(a),\\
\gamma &=& Ai(a)Bi'(a)-Ai'(a)Bi(a),\\
\kappa &=& [(-2m^*_2 eV_a)/(a \hbar^2)]^{1/3},
\end{eqnarray*}
$Ai(0)[Bi(0)]$, $Ai(a)[Bi(a)]$, $Ai'(0)[Bi'(0)]$, and
$Ai'(a)[Bi'(a)]$ are the values of Airy functions
$Ai(\eta)[Bi(\eta)]$ and their derivations with respect to $\eta$ at
$x=0$ and $x=a$, respectively. According to the stationary phase
method discussed above, the lateral shift of the ballistic electrons
transmitted through a quantum slab under applied electric field is
thus given by $s'=-d\phi'/dk_{y0}$. By numerical calculations, Fig.
\ref{fig.4} shows the lateral shifts of the ballistic electrons
transmitted though a quantum slab under an external applied electric
field $V_a$, where $V_a =0.1 V$ (solid curve), $V_a =0.01 V$ (dashed
curve), and the other physical parameters are the same as in Fig.
\ref{fig.2}. It is shown that the lateral shifts with biased voltage
can also be negative as well as positive in the same way as those in
absence of external applied electric field.

\begin{figure}[]
\scalebox{0.40}[0.40]{
  \includegraphics{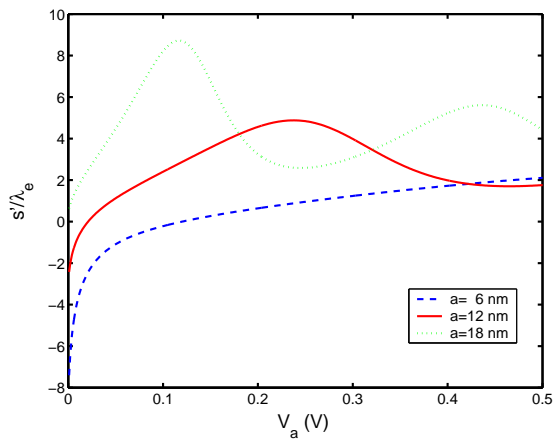}}
\caption{(Color online) Dependence of the lateral shift on the
external applied electric field, where $a=6nm$ (dashed curve),
$a=12nm$ (solid curve), $a=18nm$ (dotted curve), and the other
physical parameters are the same as in Fig. \ref{fig.2}.}
\label{fig.5}
\end{figure}
Next, we will discuss in detail on the effect of external applied
electric field on the lateral shift. Fig. \ref{fig.4} shows the
dependence of the lateral shift $s'$ on the width $a$ of the
semiconductor quantum slab under different applied electric fields.
It is clearly seen that the lateral shifts changes dramatically when
the biased voltage $V_a$ increases. In Fig. \ref{fig.5}, the lateral
shift can tuned from negative to positive under some conditions. It
is further shown that it is more feasible to control the lateral
shift by external applied electric field with increasing the width
$a$ of the semiconductor quantum slab. In Ref. \cite{Chen-PRB}, it
has been shown that there exists a large negative lateral shift in
magnetic-electric nanostructure. However, as in many previous
proposals \cite{Chen,Chen-PRB}, the positive or negative lateral
shifts of the ballistic electrons cannot be manipulated once one
chooses the quantum semiconductor structure. Here we use the applied
electric field to realize the modulation of the lateral shifts in a
fixed semiconductor quantum well. Thus, it is a very useful and
powerful way to control the spatial beam location of the ballistic
electrons, which may lead to the potential applications in quantum
electric devices, such as electric spatial modulation, beam
splitter, and electron wave vector or energy filter \cite{Kan}.

\begin{figure}[]
\scalebox{0.40}[0.40]{
  \includegraphics{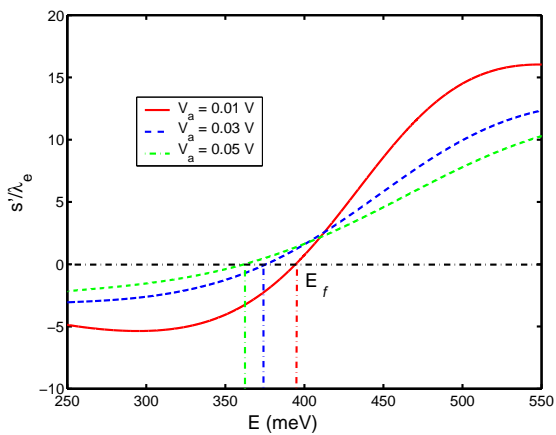}}
\caption{(Color online) Dependence of the lateral shift on the
incident energy under the different external applied electric
fields, where $a=6nm$, $V_a=0.01 V$ (solid curve), $V_a=0.03 V
$(dashed curve), $V_a=0.05 V$ (dash-dotted curve), and the other
physical parameters are the same as in Fig. \ref{fig.2}.}
\label{fig.6}
\end{figure}

Finally, we have a brief look at the applications of tunable lateral
shifts in various quantum electric devices. To this end, we
represent an example of the spatial beam modulation and energy
filter based on the voltage-tunable lateral shifts, as shown in Fig.
\ref{fig.6}, where $a=6nm$, $V_a=0.01 V$ (solid curve), $V_a=0.03 V
$(dashed curve), $V_a=0.05 V$ (dash-dotted curve), and the other
physical parameters are the same as in Fig. \ref{fig.2}. Fig.
\ref{fig.6} shows that the different incident energies of ballistic
electrons correspond to their different lateral shifts in
transmission. So the spatial beam location can be modulated by the
incident energy under the different external applied electric
fields. In addition, it is also shown in Fig. \ref{fig.6} that the
lateral shift is positive for $E>E_f$, while it is negative for
$E<E_f$. Furthermore, the threshold of energy $E_f$ corresponding to
the lateral shift $s'=0$, which is measured by the spatial location
$y=0$, can be controlled by the biased voltage $V_a$. Therefore, the
energies of incidence electrons can be easily chosen by the positive
or negative lateral shifts, which results in the energy filter. As a
result, the quantum devices such as spatial modulation, beam
splitter, and wave vector or energy filter can be realized by the
controllable negative and positive lateral shifts, because of their
dependence on incident energy and incidence angle.

\section{Summary}
In conclusion, the lateral shifts of the ballistic electrons
transmitted through a semiconductor quantum well can be positive as
well as negative. The necessary condition (\ref{restriction to
incident angle}) is advanced here for the shift to be negative. The
negative and positive lateral shifts, which is similar to but
different from the negative refraction for ballistic electron, are
analogous to those of the light beam transmitted through a
dielectric slab. This semiconductor quantum slab discussed here
behaves as the Fabry-Perot-type interferometer for ballistic
electrons in the literature \cite{Liang}. Therefore, the anomalous
lateral shifts can be understood from the destructive and
constructive interference due to the multiple reflections and
transmissions inside the quantum slab. More importantly, the lateral
shifts can be tuned from negative to positive when the external
electric field is applied in a fixed semiconductor quantum
structure. In a word, with development of semiconductor technology,
the tunable lateral shifts of ballistic electrons may lead to the
potential applications in various quantum electric devices.

\section*{Acknowledgments}
This work was supported in part by the National Natural Science
Foundation of China (60806041, 60877055), the Shanghai Rising-Star
Program (08QA14030), the Science and Technology Commission of
Shanghai Municipal (08JC14097), the Shanghai Educational Development
Foundation (2007CG52), and the Shanghai Leading Academic Discipline
Program (S30105).

\end{document}